\documentclass[prb,twocolumn,showpacs,amsmath,amssymb,letterpaper,eqsecnum,floatfix]{revtex4}

\usepackage{dcolumn,times}

\begin{document}

\def \SAVE#1{}
\def \LATER #1{}
\def \OMIT #1{}
\def \MEMO #1{{\bf #1}}


\newcommand{\la}{\langle}
\newcommand{\ra}{\rangle}
\newcommand{\fsel}{{f_{\rm sel}}}
\newcommand{\ix}{i}
\newcommand{\jy}{j}
\newcommand{\kz}{k}
\newcommand{\lw}{l}
\newcommand{\Tr}{{\rm Tr}}
\renewcommand{\AA}{{\bf A}}
\newcommand{\BB}{{\bf B}}
\newcommand{\Bbar}{{\bar{\bf B}}}
\newcommand{\evec}{{\bf e}}
\newcommand{\xhat}{{\bf \hat{x}}}
\newcommand{\yhat}{{\bf \hat{y}}}
\newcommand{\zhat}{{\bf \hat{z}}}
\newcommand{\rr}{{\bf r}}
\newcommand{\hh}{{\bf h}}
\newcommand{\MM}{{\bf M}}
\newcommand{\nn}{{\bf \hat{n}}}
\newcommand{\qq}{{\bf q}}
\newcommand{\II}{{\bf I}}
\newcommand{\JJ}{{\bf J}}
\newcommand{\PP}{{\bf P}}
\renewcommand{\SS}{{\bf S}}
\renewcommand{\tt}{{\bf t}}
\newcommand{\ssigma}{{\boldsymbol{\sigma}}}
\newcommand{\epp}{{\boldsymbol{\epsilon}}}
\newcommand{\HH}{\mathcal{H}}
\newcommand{\Jeff}{J^{\rm eff}}
\newcommand{\Jreal}{J^{\rm real}}
\newcommand{\Jpure}{J^{\rm pure}}
\newcommand{\half }{{\frac{1}{2}}}
\newcommand{\REF}{_{\rm REF}}
\newcommand{\dil}{_{\rm dil}}
\newcommand{\sel}{_{\rm sel}}
\newcommand{\HHex}{\HH_{\rm ex}}
\def \be {\begin{equation}}
\def \eeq {\end{equation}}
\def \bea {\begin{eqnarray}}
\def \eea {\end{eqnarray}}
\def \Figr #1{Fig.~\ref{#1}}
\def \q {{\bf q}}
\def \eqr #1{(\ref{#1})}
\def \Ep#1{(\ref{#1})}
\def \vol #1{{\bf #1}}

\def\cmt{$ {\rm Cd}_{1-p}{\rm Mn}_p{\rm Te}$\ }
\def\cmse{${\rm Cd}_{1-x}{\rm Mn}_i{\rm Se}$}
\def\cms{${\rm Cd}_{1-x}{\rm Mn}_i{\rm S}$}
\def\bn{{\bf N}^{\alpha\sigma}}
\def\ks{ \xi^{\alpha\sigma}(\ix)}
\def \grad {\nabla}
\def\dilsq{\langle \epsilon^2 \rangle}
\def\nabsp{|{\bf N}^{\alpha +}|^2}
\def\nabsm{|{\bf N}^{\alpha -}|^2}
\def\dref{ \langle \delta U\REF^2\rangle^{1/2} }

\title{Effective Hamiltonians for 
state selection in Heisenberg antiferromagnets}

\author
    {Brond E. Larson} 
\affiliation{Ab Initio Software, LLC,  201 Spring St.,
Lexington, MA 02421}
\author{Christopher L. Henley}
\email{clh@ccmr.cornell.edu}
\affiliation{Laboratory of Atomic and Solid State Physics, 
    Cornell University, Ithaca, New York 14853-2501, USA}

\date{\today}

\begin{abstract}
In frustrated antiferromagnets with isotropic exchange
interactions, there is typically a manifold of degenerate
classical ground states.  This degeneracy is broken by
the (free) energy of quantum or thermal fluctuations, 
or the uniform effects of bond disorder. 
We derive effective Hamiltonians to express each kind
of selection effect, in both exact forms and convenient
approximate forms.  It is argued that biquadratic terms,
representing the collinear-selecting effects of quantum
fluctuations, should be included in classical 
simulations of large-$S$ frustrated magnets at low temperatures.
\end{abstract}


\maketitle

\section {Introduction}
\label{secI}

\LATER{Check again my list of papers citing Larson-Henley}

This paper concerns antiferromagnets with isotropic
exchange couplings, i.e. the Hamiltonian has form
\be
    \HHex = -\frac{1}{2}\sum_{\ix \neq \jy}J_{\ix\jy}
   \SS_\ix\cdot\SS_\jy.  
   \label{eq:ham}
\eeq
(Such sums will always run to $N$, the number of spins.)
When the interactions in \eqr{eq:ham} are frustrated, 
the classical ground states often exhibit 
continuous {\it nontrivial} degeneracies --- beyond those 
induced by the global spin symmetry. 
In a {\it highly} frustrated antiferromagnet~\cite{ramirez},
the number of such degrees of freedom is extensive;
in other cases, such as face-centered cubic (fcc) antiferromagnets,
it is a finite-dimensional manifold.
In either case, 
such degeneracies are lifted by thermal fluctuations, quantum
fluctuations,  or quenched fluctuations (due to dilution).  
The selection of a ground state due to fluctuations
was called~\cite{Henley1987} (by a slight abuse of terminology)
{\it ordering by disorder}.~\cite{Villain}
The purpose of this paper is to provide effective Hamiltonians,
in convenient form, which represent the degeneracy-breaking
selection (free) energy.

A ground state selection calculation for a continuous spin system
was performed first by Shender, who showed that in the
two-sublattice garnet 
${\rm Mn}_3{\rm Cr}_2{\rm Ge}_3{\rm O}_{12}$ 
quantum fluctuations favor a collinear spin
configuration.~\cite{Shender1} The selection causes an effective
``anisotropy'' of one sublattice relative to the other, which
opens a gap in some soft spin wave modes observable in neutron
scattering.~\cite{Shender2,Shender3} 
Others have calculated a similar effect for the type I FCC
antiferromagnet $\gamma$-Mn$_p$Fe$_{1-p}$~\cite{Oguchi,LongandYeung,Long1989} .
Rastelli and co-workers investigated
thermal and quantum fluctuations in the frustrated
rhombohedral lattice (describing $\beta-O_2$), finding again that
quantum fluctuations select collinear 
ground states~\cite{Rastelligroup}.
\LATER{Add references: ``and similarly in frustrated tetragonal lattices like
cuprates: Harris, Aharony...''.}
The collinearity bias was decisive between different possible
fcc orderings in Cu nuclear spins~\cite{Lindgard,oja}.
Quantum fluctuations were also found to decide the
ground state in Sr$_2$Cu$_3$O$_4$Cl$_2$~\cite{Kim99}.

Whereas thermal or quantum fluctuations select 
{\it collinear} states~\cite{Shender1,HenleyPRL}
quenched fluctuations due to dilution select {\it noncollinear}
states~\cite{Henley1987,HenleyPRL}, in cases where the ground state manifold
is finite dimensional.  (In {\it highly} frustrated magnets, meaning
the degrees of freedom are extensive, the many zero-energy
excitations ``screen'' defects and dilution does not necessarily
favor a global noncollinear state~\cite{shender-cherepanov,Henley-HFM2000}.)
The competing selection effects can give a rich phase
diagram as a function of temperature and dilution, 
as simulated in the $J_1$-$J_2$ square lattice 
antiferromagnet~\cite{HenleyPRL,Prakash,Chandra1989,FernandezXY}.
In particular, making use of the effective Hamiltonians presented 
in this paper, we studied the fcc type III antiferromagnet with 
dilution and Dzyaloshinskii-Moriya anisotropic exchange~\cite{larson-henley};
this represents the antiferromagnetic dilute magnetic 
semiconductor \cmt~ for $x\to 1$, a material
realized using molecular-beam  epitaxy~\cite{giebultowicz-sim,Samarth}.
\LATER{did I cite Samarth-Giebultowicz expts c 1990?}

For all three kinds of selection -- thermal, quantum, and dilution 
-- the selection is conveniently expressed by adding 
biquadratic exchange terms to the Hamiltonian,
\be
   \HH_{\rm biq}=-\half \sum_{\ix\jy}
   K_{\ix\jy} \Big(\SS_\ix\cdot \SS_\jy \Big)^2 .
   \label{eq:biquad}  
\eeq
Such a form is quite familiar for selection by quantum 
fluctuations in the ``independent sublattices''
case, i.e. the antiferromagnet consists of 
two (pr more) sublattices with a strong, unfrustrated  intra-lattice
intereaction  $J_1$, and a weak, frustrating inter-lattice
interaction $J_2$~\cite{Shender1,Shender2,HenleyPRL,Yildirim}:
Eq.~\eqr{eq:biquad} follows directly from perturbation in 
$J_1/J_2$, with inter-sublattice $K_{\ix\jy} = O(J_1^2/J_2)$.
However, there are many other cases in which the degeneracy
freedom requires correlated rotations in all the sublattices.
This paper lays out the framework in which \eqr{eq:biquad}
can be justified (approximately) in that broader class of 
frustrated systems.

In Sec.~\ref{sec:classical}
we present a microscopic calculation of the free
energy in the presence of quenched disorder and thermal
fluctuations, performed to lowest nontrivial order in spin
deviations.  The combined selection effects are
extracted in a new way by performing a constrained
integration over the fluctuations in reciprocal space.  
Then, in Sec.~\ref{sec:effham} we express the results
in the phenomenological form of an effective biquadratic
exchange term.
In Sec.~\ref{sec:quantum}, quantum
fluctuations are shown to give a similar form,
using an approximation of the spin-wave approximation.
Appendices ~\ref{app:alt-dilution} and \ref{app:magnetoel}
give, respectively, an alternate path to the dilution
effective Hamiltonian, and a calculation of magnetoelastic
effects (which reduce to a similar quartic effective Hamiltonian, 
and thus give selection effects similar to those of thermal or
quantum fluctuations).
The Discussion (Sec.~\ref{sec:discussion})
proposes applications for the effective Hamiltonian in
working out phase diagrams in simulations of magnetic systems.

\section{Thermal Fluctuations and Dilution}
\label{sec:classical}
 
In this section we show that thermal fluctuations favor collinear
states while dilution favors the least collinear states.  This
competition is a general feature of all uniformly frustrated
vector spin systems studied so far and has a simple
explanation~\cite{Henley1987,HenleyPRL}.  
It is familiar
that spins in a vector antiferromagnet will orient transverse to
an applied magnetic field, because they can thereby gain energy
by relaxing towards it.  The {\it effective} local fields generated
by thermal fluctuations are transverse to the fluctuating spins,
and the free energy will be lower when other spins are transverse
to these effective fields.  The coupling, and the entropy, is
therefore maximized when all spins are {\it collinear}~\cite{Shender1}.

By contrast, dilution (the removal of
spins) generates effective fields parallel to the removed spins
(or other spins of the same sublattice).  In this case the energy
is minimized for the {\it least} collinear states, because most
spins are again transverse to the effective field acting on them.
The above qualitative arguments will be confirmed by 
perturbative spin-mode calculations.  

In the rest of this section, we first set up the artificial 
form of dilution; then, representing a delta function by 
auxiliary variables, we sidestep the awkwardnesses 
expanding about a general nonperiodic, noncollinear classical
ground state (Subsec.~\ref{sec:classical-formal}).
This result can be put in the form of a biquadratic 
effective Hamiltonian (Sec.~\ref{sec:effham}).

\subsection{Formal Derivation of Selection Free Energy}
\label{sec:classical-formal}


For this section, the spins in \eqr{eq:ham} are taken 
to be classical unit vectors with $n$ components.  
We compute the free energy by assuming the fluctuations about the
ground state are small, 
\be
    \SS_\ix=\SS^{(0)}{}_\ix + \ssigma_\ix
    \ \ ,\ \ \ \ \ \  \SS^{(0)}{}_\ix\cdot\ssigma_\ix=0~~,
\eeq
and $| \sigma | \ll 1$.
Further, we will consider ``modulation'' disorder where the
exchange between sites $\ix$ and $\jy$ is increased or
decreased by the independent random variables $\epsilon_i$.
The actual couplings are taken to be
\be
   \Jreal_{ij}=  J_{ij}(1-\epsilon_i) (1-\epsilon_j)~~, 
   \label{eq:modis}  
\eeq
with $\langle \epsilon_i \rangle = 0$.  
This approximation~\cite{fncu} 
lets us control the size of the
deviations $\ssigma_i$ by taking the variance of the random
modulation to be a small parameter: 
$\langle \epsilon_i{}^2 \rangle \equiv \dilsq \ll 1$.  
We will take $J_{\ix\jy}^{\prime}=\Jpure_{\ix\jy}$ for
modulation disorder, where $\Jpure_{\ix\jy}$ is the 
energy of a bond in the pure system.

True site dilution means 
$\Jreal_{ij}=\Jpure_{ij}\eta_\ix \eta_iy$, where the
random variable $\eta_\ix$ is 1 (site occupied) with probability $p$ and zero
with probability zero. (See Appendix~\ref{app:alt-dilution}.)
This takes the form of Eq.~\eqr{eq:modis}, with 
$\langle \epsilon_i\rangle=0$, if we take $\epsilon_i \equiv 1-\eta_i/p$.
This then forces $J_{ij} = p^2 \Jpure_{ij}$.

For the ``independent lattices'' case e.g.
the type II bcc~\cite{Shender1,Shender2},
the type II FCC ordering, the frustrated square lattice model~\cite{HenleyPRL}, 
or a body-centered tetragonal antiferromagnet~\cite{Yildirim},
the ratio $J_1/J_2$ can be a small parameter for achieving
this same control in the limit of large $| J_2 | $,
because the noncollinear part of the effective field of a missing
spin becomes small compared to the collinear part.  In 
Appendix~\ref{app:alt-dilution}
it is shown that the model of \eqr{eq:modis} and true
dilution are equivalent to lowest order in the spin fluctuations.
Monte Carlo results for true dilution are consistent with the
results of this theory for $p \lesssim 1$.~\cite{Prakash,larson-henley}

The ground state condition is that $\SS^{(0)}_i = \hh^{(0)}_i/|\hh^{(0)}_i|$
for all $i$, i.e. every spin is aligned with its ``local field''
  \be
     \hh^{(0)}_i \equiv \sum_j J_{\ix,\jy} \SS^{(0)}_j~~.
  \label{eq:h0}
  \eeq
We assume $|\hh^{(0)}_i|\equiv h_0$, the same at every site
(subsequent manipulations depend on this.)
The assumption is valid on most cases that the
sites are symmetry-equivalent,
even in highly frustrated systems 
(e.g. pyrochlore lattice)
where a generic ground state is not at all periodic.
Substituting \eqr{eq:h0} in \eqr{eq:ham}
shows $E_0=h_0/2$, so $h_0$ must be the same in
every ground state, too.

Expanding in small deviations,
the Hamiltonian \Ep{eq:ham} becomes 
a quadratic form
\bea
    \HH_{\rm quad}-NE_0 &=&\sum_{i}\hh_i
  \cdot \SS^{(0)}{}_i +\sum_{i} \hh_i \cdot \ssigma_i  \nonumber\\
   &&\quad + \half \sum_{ij} A(\ix,\jy) 
   \ssigma_i \cdot \ssigma_j 
   + {\cal O}(\epsilon^3)
   \label{eq:hamdil} 
\eea
Here $E_0$ is the ground state energy per spin of the pure system,
and 
   \be
     \hh_i \equiv \sum_j J_{\ix,\jy}\epsilon_j\SS^{(0)}_{j}~~.
   \label{eq:hrandef}  
   \eeq
Also, 
   \be
    A_{\ix,\jy} \equiv
    -J_{\ix,\jy} + h_0\delta_{ij}~~. 
   \label{eq:adef} 
\eeq

The first term on the right hand side of \eqr{eq:hamdil}
is independent of $\ssigma_i$ and vanishes when
the configurational averages are taken.  We ignore the last term
since it is higher order in $\epsilon $.

\subsubsection{Evaluation via auxiliary representation of constraints}

We now have a
Hamiltonian which is purely quadratic and apparently trivial.
However, as always, each spin really has only $n-1$ degrees of
freedom since its length is fixed.  
When $n>2$, there is an arbitrariness in the choice of 
basis vectors for the local transverse subspace, 
a gauge freedom.  
It is cumbersome to fix this gauge by a choice of
local frames~\cite{WalkerandWalstedt}.
Instead, following Ref.~\onlinecite{Henley1987}
we shall 
implement the constraints so as to avoid 
introducing an explicit local frame for each spin,
using integral identities to maintain a manifestly 
rotation-invariant (gauge-invariant) form.

For convenience, a matrix notation will be used where the matrix
indices specify both position ($\ix$) and vector component in
spin space ($\mu $).  The unit length constraint becomes
\be
    \PP\cdot \ssigma=0.
\eeq
Here $\PP$ is the nonsquare $(N \times Nn)$ matrix whose components are 
\be
    P_{\ix,\ix^{\prime}\mu^{\prime}}=\delta_{\ix,\ix^{\prime}}
  S^{(0)}{}_{\ix^{\prime}{\mu}^{\prime}}.  
\eeq
Since $\SS^{(0)}_\ix$ is a unit vector, 
    \be
                 \PP\PP^T = \II_{N\times N},
    \label{eq:PPT}
    \eeq
where $\II$ is the identity matrix, and we (sometimes)
add a subscript to clarify a matrix's dimension.
(However, notice $\PP^T \PP \neq \II_{Nn\times Nn}$.)
The definition of $A_{\ix,\jy}$ [Eq.~\eqr{eq:adef}] becomes
\be
    \AA=-\JJ + h_0\II.
    \label{eq:amatdef} 
\eeq
In \eqr{eq:amatdef}, the matrices $\AA$ and $\JJ$ are 
$N\times N$, but from here
till the end of the section they are extended to be $Nn\times Nn$
matrices (by taking the direct product with $\II_{n\times n}$,
acting on the spin indices).
Also, $\hh$ [from \eqr{eq:hrandef}] becomes an $Nn$-component vector
in this notation.

Our goal, the partition function for Hamiltonian \eqr{eq:hamdil},
is 
\be
   {Z=\int\Big[\prod_{\ix\mu}d\ssigma_{\ix\mu}\Big]
    \Big[\prod_\jy \delta (\PP\cdot \ssigma)\Big] 
     \exp \big(-\half \beta\ssigma^T 
     \AA \ssigma + \beta \hh^T\ssigma\big).} 
   \label{eq:part} 
\eeq
By representing the delta function of
the constraints in terms of a functional integral of 
$\exp(i{\tt \PP\ssigma})$ 
over auxiliary
variables $ \{ t_\ix \}$ and completing the square in the resulting
exponential, the $d\ssigma_{\ix\mu}$ integral can be done~\cite{Henley1987},
yielding
\bea
   Z=Z_1\int\Big[\prod_\ix \frac{dt_\ix}{2\pi}  \Big]
   &&\exp \Big(-\half ( \hh^T +
   i \beta^{-1} \tt^T\cdot \PP)\AA^{-1} \nonumber \\
       && (\hh +i\beta^{-1}\PP^T \cdot \tt)\Big).
   \label{eq:partb} 
\eea
Here
\bea
    Z_1&\equiv& \int\Big[\prod_{\ix\mu}d\ssigma_{\ix\mu}\Big]
\exp\Big(-\half \beta \ssigma^T \AA  \ssigma\Big) \nonumber \\
   &=& \Big(\frac{2\pi}{\beta}\Big)^{\frac{Nn}{2}}  
   \big(\det A\big)^{-\half}.
\eea

To make further progress, let us define the ($N \times N$)
matrix of the quadratic coefficients in \eqr{eq:partb}:
  \be 
    \BB\equiv \PP\AA^{-1}\PP^T,
    \label{eq:Bdef}
  \eeq
\LATER{DO WE REALLY NEED THIS OR MERGE INTO NEXT SECTION:
(Thus its components are
   $B_{\ix,\jy} \equiv \SS^{(0)}{}_\ix\cdot \SS^{(0)}{}_\jy
   A^{-1}_{\ix,\jy}$. )
}
All our results will be expressed in terms of $\BB$.
Since $\BB$ is nonsingular, we can evaluate 
\eqr{eq:partb} using a second completion of squares~\cite{FN-PPnotsquare}.

Thus,
\be
   Z=Z_1~ Z_2 ~ e^{-\beta\Delta F},
   \label{eq:partab} 
\eeq
 where
\be
   Z_2 \equiv \int\Big[\prod_\ix \frac{dt_\ix}{2\pi} \Big]
   e^{-\half \beta^{-1}\tt^T \tt}
      =[2\pi\beta]^{\frac{N}{2}}(\det \BB)^{-\half }, 
\eeq
and 
    \be
       \Delta F \equiv -\half \hh^T \Big(\AA^{-1} - \AA^{-1}\PP^T
       \BB^{-1}\PP\AA^{-1} \Big)\hh.
    \label{eq:DeltaF}
    \eeq

\subsubsection{Free energy result and disorder average}

Thus the exact free energy is
\bea
  \label{eq:free}
   F&=&-\beta^{-1} \Big[N\big(1+{\frac{n}{2}}\big )\ln(2\pi)+
                 N\big(1-{\frac{n}{2}}\big )\ln \beta
            -\half \ln(\det\AA) \Big] \nonumber \\
   &+&\beta^{-1}\half \ln(\det\BB) +
   \Delta F -\hh^T
   \cdot \SS^{(0)} \nonumber \\
   &\equiv& F_0 + F_T + F\dil + F\REF
     \equiv F_0 + F\sel,
\eea
where we have reintroduced the linear term
in $\hh$ from \eqr{eq:hamdil}.  In this equation, the term
in brackets ($F_0)$ is independent of the ground state, 
hence irrelevant for state selection (though possibly important
in comparing to the free energy of competing orders).
The remaining three terms constitute the {\it
selection free energy}: respectively,  
thermal selection, dilution selection term,  and finally 
$F\REF$ linear in the
effective fields caused by dilution. 
The disorder average of $F\REF$ is clearly zero, 
but such terms give ``random exchange fields'', 
so called as they (may) act like a random field on the discrete
order parameter remaining after selection
\cite{Henley1987,fernandez,FernandezXY,larson-henley}.

Next we average $F\sel$ over disorder.  
Recall from \eqr{eq:hrandef}
that $\hh= \JJ\PP^T\epp$ 
and 
$\langle \epsilon_\ix \epsilon_\jy\rangle = \delta _{\ix,\jy}\dilsq $ 
allows us to average $\Delta F \equiv F\dil$ 
[given by \eqr{eq:DeltaF}]:
\be
   \langle F\dil \rangle = -\half \dilsq \Tr\Big[
    \PP \JJ \big(
      \AA^{-1} - \AA^{-1} \PP^T\BB^{-1} \PP \AA^{-1}\big)
        \JJ \PP^T \big]
   \label{eq:avdelf} 
\eeq
Substituting $\JJ \to h_0 \II -\AA$ from \eqr{eq:amatdef}, 
Eq.~\eqr{eq:avdelf} expands into zero, first, and second
order terms in $h_0$.  By repeated use of 
\eqr{eq:PPT} and \eqr{eq:Bdef}, the contributions of the
two terms in parentheses inside \eqr{eq:avdelf} 
are seen to cancel in both the $O(h_0)$ and $O(h_0^2)$ 
parts of the expansion.  
The result is
\be
   \langle F\dil \rangle = -\half \dilsq 
       \Tr \big(\PP\AA\PP^T - \BB^{-1}\big)
   \label{eq:avdil}  
\eeq
Now $\Tr (\PP\AA\PP^T) = \Tr(\AA_{N\times N})=  Nh_0$,
which is the same for all ground states.  
Gathering terms, the averaged selection free energy is
\be
   { \langle F\sel \rangle = 
   \half k_BT \; \Tr \big(\ln\BB \big)
   + \half \dilsq \Tr\big(\BB^{-1}\big)-\half \dilsq Nh_0,} 
   \label{eq:fsel} 
\eeq
where we have used $\ln(\det \BB) = \Tr (\ln \BB)$.

It can be shown that \eqr{eq:avdil} vanishes for
collinear states and is {\it negative} semidefinite 
(see Appendix~\ref{app:alt-dilution}).
Therefore dilution must favor noncollinear states.

\subsection{Effective Hamiltonian for Selection}
\label{sec:effham}

\LATER{Mention:
The effective Hamiltonian 
is defined only for states in the ground state
manifold;  it does not apply to fluctuations out
of that manifold, which we integrated out in Sec.~\ref{sec:classical-formal}.}

In this section we re-express the classical selection 
(free) energy \eqr{eq:fsel} 
in the form of an effective Hamiltonian.
In this form, it is easy to join the selection
terms with other perturbations that also tend to
select states, such as anisotropies
or magnetoelastic couplings.
The selection term takes the biquadratic form \eqr{eq:biquad}.
for all exchange-coupled frustrated vector magnets.

In the $J_1$-$J_2$ antiferromagnets on the square lattice 
[$(\pi,0)$ order], the fcc lattice (type II order) 
or the bcc lattice (bcc type II order),
the ground states consist of sublattices which can 
rotate independently of each other 
(provided order-by-disorder is neglected!).
These sublattices are connected only by $J_1$ interactions, 
so $J_1/J_2$ can be used as a small parameter.
Indeed, the lowest-order term in the $J_1/J_2$
expansion does have the biquadratic form 
\Ep{eq:biquad}  for the $J_1$-$J_2$ XY model 
on the square lattice~\cite{HenleyPRL} or for quantum fluctuations in the 
large-$S$ Heisenberg model on the bcc~\cite{Shender1}.
However, for many other cases (e.g. fcc type I or type III),
the degeneracies entail
correlated changes in different sublattices, 
so there is no natural small parameter.

We deploy a different trick, which has no guaranteed
small parameter.  It is based on the central
role played by the matrix $\BB$ in the previous subsection.
Recall \eqr{eq:fsel} gave the selection free energy in terms of
$\ln\det\BB$ and ${\rm Tr}\BB^{-1}$. 
The matrix $\BB$ was defined in real space by \eqr{eq:Bdef}
as 
    \be
       \BB_{ij}=\SS_i\cdot \SS_j G_{ij}
    \eeq
where 
   \be
       G_{ij}\equiv \big(\AA^{-1}\big)_{ij} .
   \eeq
is a Green's function.
(For this section, we drop the superscript in $\SS^{(0)}_i$; 
it is understood that $\SS_i$ is a mean direction.)
Although $G_{ij}$ has the translational invariance of
the lattice, it is {\it not} simply a function of $\rr_i-\rr_j$
if the sites do not form a non-Bravais lattice 
(e.g. honeycomb, garnet, or pyrochlore lattices).
Since the $\AA$ matrix is a discrete approximation
of the Laplacian, $G_{ij}$ is likely to behave 
at long distances as $1/r_{ij}$ (in $d=3$) 
or as $\ln r_{ij}$ (in $d=2$).

Now for the trick:
let $\Bbar$ be the average of $\BB$ over the whole ground state
ensemble, 
\be
    \Bbar_{ij}=\langle \SS_i\cdot \SS_j \rangle_s G_{ij}, 
   \label{eq:bave} 
\eeq
where the angle brackets refer to an
unweighted average over all ground states.~\cite{fnweight}
(Notice $\Bbar$ has the full symmetry of the {\it lattice}, 
higher symmetry than $\BB$ has.)
We shall formally consider 
\be
   \delta \BB \equiv \BB-\Bbar
   \label{eq:bdel} 
\eeq
to be our small parameter.
Thus
   \be
     \delta \BB_{ij} = G_{ij} 
     \big(\SS_i\cdot \SS_j - \langle \SS_i\cdot \SS_j \rangle_s\big) 
   \eeq

Then from \eqr{eq:fsel} 
by expanding the logarithm of $\BB$ (for thermal selection) 
or its inverse (for dilution selection) in powers of $\delta \BB$, 
we arrive at 
\be
    \langle F\sel \rangle \approx [\fsel^{(0)} + 
   \fsel^{(1)} + \fsel^{(2)}],
   \label{eq:fexpansion} 
\eeq
where $\fsel^{(m)}$ is the $O\big((\delta \BB) ^m \big)$ term.
In \eqr{eq:fexpansion}, 
\bea
   \fsel^{(0)}&=& \half k_BT\ln(\det\Bbar) + \half \dilsq 
   \Tr (\Bbar^{-1}), \\
    \fsel^{(1)}&=& \half \sum_{i,j} \Jeff_{ij} \SS_i \cdot \SS_j,
   \eea
where
  \be
    \Jeff_{ij} \equiv 
     2k_BT\sum_{i,j}(\Bbar^{-1})_{ij}G_{ji}
    -2\dilsq \sum_{i,j} (\Bbar^{-2})_{ij}G_{ji} ;
     \eeq
and 
   \be
   f^{(2)}=-\frac{1}{4}  \sum_{i,j,k,l}  K_{ij,kl}
       (\SS_j\cdot \SS_k) ( \SS_l\cdot \SS_i),
   \eeq
with
  \bea
   K_{jk,li}&\equiv& G_{jk} G_{li} \Big[
    k_BT (\Bbar^{-1})_{ij} (\Bbar^{-1})_{kl} \nonumber\\
   &+& \dilsq (\Bbar^{-2})_{ij}
   (\Bbar^{-1})_{kl} \Big]_{\rm symm}
  \label{eq:biquadexp} 
   \eea
Here ``symm'' means the expressions are to be symmetrized
over all distinct ways of pairing indices, such that
each pair has one from $jk$ and one from $li$
(two or four ways, respectively, in the two terms).
Each term includes contributions from both $F_T$ and $F\dil$.
 
\SAVE{The original text had an errors in \eqr{eq:biquadexp}.
The last term had $\Bbar^{-1}$, where it SHOULD have had $\Bbar^{-2}$.}

Clearly $f^{(0)}$ does not cause selection, since it produces
only constants dependent on $T$ and $\dilsq $.  
As for the $f^{(1)}$ term of \eqr{eq:biquadexp},  it takes the
form of a renormalization of the ordinary exchange couplings 
$J_{\ix\jy} \to J_{\ix\jy}  + \delta J_{\ix\jy}$, where
  \be
   \delta J_{\ix\jy} \equiv 
    2k_BT(\Bbar^{-1})_{\ix\jy}G_{\jy\ix}
   -2\dilsq \sum_{i,j}
   (\Bbar^{-2})_{\ix\jy}G_{\jy\ix}.
  \eeq
(Here $\delta J_{\ix\jy}$ has the full lattice symmetry,
since $\Bbar$ and $G_{\ix\jy}$ do.)
Within the ground state manifold, $f^{(1)}$ is 
normally constant (assuming the degeneracy is
generic by symmetry, and not due to a particular
ratio of the couplings).  Nevertheless, the renormalization
will shift the phase boundary between the kind of order
in question and a competing spin order.

When can we simplify $f^{(2)}$?
Commonly, and certainly when the sites
form a Bravais lattice~\cite{AFM-LT-59,Kaplan06}
one can write any classical ground state 
as a linear combination
of the (degenerate) optimal eigenvectors of the $J_{\ix\jy}$
matrix, the so-called Luttinger-Tisza construction~\cite{FN-LT-bave}.
The (vector valued) coefficients play the role of 
order parameter components $\{ {\bf N}_\alpha \}$.
Obviously, $f^{(2)}$ will be quartic in 
$\{ {\bf N}_\alpha \}$.  If group theory tells us there are
$m$ independent quartic combinations -- 
$m=2$ for the type III fcc antiferromagnet~\cite{larson-henley} --
then the parameter space of possible $f^{(2)}$ functions
is only $m$ dimensional.  We could exactly reproduce these
functions using \eqr{eq:biquad} out to the $m$-th nearest neighbor.


\subsection{Real-space perturbation approach}

There is an alternative, cruder path to the
biquadratic form, in the case of classical thermal
fluctuations~\cite{canals-zhitomirsky}.
It is included here because it is fully analogous
to the approach we use for quantum selection in
Sec.~\ref{sec:quantum}.

The idea in this approach is to treat the
site-diagonal part of \eqr{eq:hamdil}
as the zero-order part, and to pretend the inter-site 
terms are a small perturbation $\HH'$.  (In fact the site-diagonal
and inter-site coefficients are manifestly the same order
in the $J_{\ix,\jy}$'s. However, as elaborated
in Sec.~\ref{sec:quantum-real},  we may consider
$1/z$ to be a sort of small parameter, where $z$ is
the coordination number.)

We continue to assume that the local field
$h_0$ is the same on every site.  Thus our zero-order
Hamiltonian is quite trivial  and tractable,
and the perturbation free energy is derived from
the cumulant expansion to second order:
   \be
     \HH_{\rm eff}  = -\la {\HH'} ^2 \ra _0 / 2 T,
   \eeq
where $\la ... \ra _0$ means the expectation in
the ensemble with just $h_0$.
(I omit the terms in $\la \HH'\ra_0 $,   
since $\la \HH'\ra_0 =0$.) 

By taking expectations in this trivial ensemble,
one sidesteps most technical annoyances due to the 
different orientations of spins, which 
necessitated the auxiliary integration in
\eqr{eq:partb}.
We simply have
   \be
       \la \sigma_{\ix\mu} \sigma _{\ix\nu} \ra_ 0  =
       \Big( \delta _{\mu\nu} - n_{\ix\mu} n_{\ix\nu}\Big) \frac{k_B T}{h_0}
   \eeq
where $\nn_i\equiv \SS_i$. (This notation for the unit spins
is consistent with Sec.~\ref{sec:quantum}.)
Then 
    \be
 \la {\HH'}^2\ra _0 = \sum _{\ix\jy} 
      \Big\la \big( J_{\ix\jy} 
       \ssigma_\ix \cdot \ssigma_\jy)^2 \Big\ra _0
   \label{eq:Hsq-exp}
   \eeq
Terms mixing two different pairs vanish since
$\la \ssigma_\ix \ra_0 =0$.
In \eqr{eq:Hsq-exp}, 
    \bea
       \big\la (\ssigma_\ix \cdot \ssigma_\jy)^2 \big\ra _0
     &=& \sum _{\mu\nu} \la \sigma_{\ix\mu} \sigma_{\jy\mu} 
     \sigma_{\ix\nu} \sigma_{\jy\nu}  \ra _0  \\
        &=& -\sum _{\mu\nu} 
       \big( \delta _{\mu\nu} - n_{\ix\mu} n_{\ix\nu}\big)  \\
       &&\quad \times \big( \delta _{\mu\nu} - n_{\jy\mu} n_{\jy}\big) 
           \Big(\frac{k_B T}{h_0}\Big)^2  \nonumber \\
      &=& \Big [ 1 + (\nn_\ix\cdot \nn_\jy)^2 \Big]
             \Big(\frac{k_BT}{h_0}\Big)^2   .
   \eea
Putting  it all together, we get the
form \eqr{eq:biquad} with 
    \be
         K_{\ix,\jy} = k_B T \frac{J_{\ix,\jy}^2}{h_0^2}
     \eeq
That is exactly what we would get from \eqr{eq:biquadexp} if
we took $A_{\ix\jy}  = h_0 \delta_{\ix\jy}$ in place
of \eqr{eq:amatdef}, which would be consistent with our
pretence that ${\bf J}$ is small compared to $h_0$.

\section{Harmonic Quantum Fluctuations}
\label{sec:quantum}

A similar effective Hamiltonian emerges from an 
approximate treatment of the selection
due to quantum spin fluctuations within the spin-wave
approximation.  The basic idea of the method was used by
Long~\cite{Long1989} 
for calculating quantum selection in type I
FCC antiferromagnets.  We reformulate this approach to be
explicitly independent of the ``gauge'' arbitrariness in choosing
local frames (Sec. III B), and show that quantum fluctuations
always favor collinear states.

\subsection{Set-up for harmonic Hamiltonian}

We start from a classical ground state spin configuration, 
defined by a set of classical directions $\{ \nn_i \}$, and
set up the Holstein-Primakoff representation in the 
standard fashion (for noncollinear spins~\cite{WalkerandWalstedt}).
In contrast with Sec.~\ref{sec:classical}, from here on
$\SS$ denotes the spin operator and is not normalized to
unit length.
A local orthonormal triad $(\xhat_i,\yhat_i,\zhat_i$) such that
$\zhat_i\equiv \nn_i$, and we quantize along the $\zhat_i$ axis;
we let a bar distinguish spins written in this local basis.
Holstein-Primakoff bosons are introduced as usual:
    \begin{subequations}
    \label{eq:hpdef}
    \be
    \bar{S}_{iz}=S-a^{\dagger}_i a_i ,
    \eeq
    \be
        \bar{S}_{i-}=a^{\dagger}_i\sqrt{2S -a^{\dagger}_ia_i} 
    \eeq
    \end{subequations}
where $\bar{S}_{i+}=\bar{S}_{i-}^{\dagger}$.
(We restored the explicit powers of $S$ that were omitted
in the classical discussion of Sec.~\ref{sec:classical}.)
Of the Taylor series for the square roots,
giving the semiclassical expansion in $1/S$, we
just  need the lowest order approximation, 
    $\bar{S}_{i-} \approx \sqrt{2S} a^{\dagger}_i$.
That means
   \be
    \SS_i \approx \bar{S}_{iz} \zhat + \frac{\sqrt{S}}{2} \Big(
           \evec_+ a^{\dagger}_i + \evec_- a_i \Big)
   \label{eq:hpdef-linear}
   \eeq
Here  $\evec_\pm \equiv ({\bf \hat{x}}\pm i{\bf \hat{y}})/\sqrt{2}$.
Next \eqr{eq:hpdef-linear} must be substituted into 
the spin Hamiltonian \eqr{eq:ham}.
The zeroeth order term (in $\{ a^\dagger_i\}$ and $\{ a_i\}$)
is the classical ground state energy, while the first order term 
vanishes since the ground state is classically stable 
(each $\nn_i$ is lined up with the local field $\hh_i$).
So we only treat the second order, i. e. harmonic  terms. 
The resulting quadratic Hamiltonian is a quantum analog of 
\eqr{eq:hamdil}:
   \bea
      H_{\rm harm} &=& 
           {\rm  const} -   \sum _i h_i (S-{a^\dagger}_i a_i) \nonumber\\
      &+& \half \sum _{ij} J_{ij}
     \Big(\evec_+ a^{\dagger}_i + \evec_- a_i \Big)
     \Big(\evec_+ a^{\dagger}_j + \evec_- a_j \Big)
   \label{eq:Hharm}
   \eea
The usual course is to (Bogoliubov) diagonalize \eqr{eq:Hharm},
giving the harmonic spin wave modes and frequencies $\{ \omega_m \}$.
Then the harmonic zero-point energy is
   \be
     E_{\rm harm}\big(\{\nn_i\}\big) = \sum _m \half \hbar \omega_m .
   \label{eq:Eharm}
   \eeq
Since the spin-wave spectrum is usually different for symmetry-
unrelated states, \eqr{eq:Eharm} typically breaks the degeneracy
with the energy scale $O(JS)$ (down by $1/S$ compared to the total
classical energy.)  As in the previous section, 
we desire simple explicit expressions for the dependence on $\nn_i$
that is implicit in \eqr{eq:Eharm}, i.e.,  an effective Hamiltonian.

To implement this for a frustrated or highly frustrated antiferromagnet requires
\begin{itemize}
\item[]
(i) The assumption -- not always valid -- of a periodic state;
in any case, the magnetic unit cell is typically large
leading to large matrices;
\item[]
(ii) The choice of an explicit frame for  each site,
normal to its spin, as the basis for possible spin deviations, 
with its gauge arbitrariness~\cite{WalkerandWalstedt};
\item[]
(iii) the calculation should be repeated for every one of its
({\it continuum} of) inequivalent ground states.
\end{itemize}
As in the classical selection calculation, all this makes the calculation 
cumbersome and sometimes impossible.

\subsection{Real-space perturbation approach}
\label{sec:quantum-real}

To get an approximate answer of simple form, 
we shall decline to diagonalize the
Hamiltonian, and instead treat the inter-spin terms 
in \eqr{eq:Hharm} as if they were small perturbations:
that is, $\HH_0$ is defined to be the first term, and
$\HH'$ is the second term.  
(This was pointed out by Long~\cite{Long1989}.)

The ground state $|0\rangle$ of $\HH_0$ 
is unique and obviously $\bar{S}_{iz}=S$ 
for all sites;  the first nontrivial term must come from second-order
perturbation theory.  The excited-states in $\HH'|0\rangle$ have
two spin flips on coupled sites $(i,j)$, with an excitation cost
$S(h_i + h_j)$, as created by the term with two boson creations.
(In most frustrated systems, $h_i =h_0$ the same on all sites,
allowing for simplifications.)
Thus, in second order perturbation theory, 
we get the correction~\cite{Long1989}
\be
    \delta E = -{\frac{S^2}{2}}\sum_{i,j}\frac{J_{ij}^2}{h_i+h_j}
| \evec_{i+} \cdot \evec_{j+}| ^2. 
   \label{eq:qmsel}  
\eeq
Now we take advantage of the fact that $\{ \xhat _i, \yhat _i, 
\zhat_i\}$ are rows of a (proper) orthogonal matrix: 
\begin{subequations}
  \bea
     x_{i\lambda} x_{i\mu} + y_{i\lambda} y_{i\mu} + z_{i\lambda} z_{i\mu} 
     &=& \delta _{\lambda \mu} ; \\
     x_{i\lambda} y_{i\mu} - y_{i\lambda} x_{i\mu} &=&
          \epsilon _{\lambda\mu\nu} z_{i\nu}
  \eea
\end{subequations}
Hence,  assuming the summation convention,
\bea
   |\evec_{+} \cdot \evec_{+}'| ^2 &=&
     (x_\lambda-iy_\lambda) (x'_\lambda-iy'_\lambda)
    (x_\mu-iy_\mu) (x'_\mu-iy'_\mu) 
      \nonumber \\
   &=& \big(\delta _{\lambda\mu} - z_\lambda z_\mu+ 
         i\epsilon_{\lambda\mu\nu} z_\nu\big)
    \big(\delta _{\lambda\mu} - z'_\lambda z'_\mu+ 
         i\epsilon_{\lambda\mu\nu} z'_\nu\big)
      \nonumber \\
   &=& 
        \big(1 -  \zhat\cdot \zhat'\big)^2
     \eea
Substituting into \eqr{eq:qmsel}, we get the main result
   \be
    \delta E = -{\frac{S^2}{2}}\sum_{i,j}\frac{J_{ij}^2}{h_i+h_j}
      \Big( 1 - \nn_i\cdot \nn_j \Big) ^2
   \label{eq:spwave-result} 
   \eeq

Is there a small parameter in this expansion?
The small parameter which justifies keeping just
the harmonic-order term is $1/S$, as always in a
Holstein-Primakoff expansion, and so the selection 
term is of order $1/S$ relative to the classical energy.
expansion at {\it harmonic} order.
But we made a second, more brutal approximation 
of the harmonic-order in treating the second term
in \eqr{eq:Hharm} as a perturbation: both terms
there are of order $S$ and of course $h_i \equiv S |\sum _j J_{ij} \nn_j|$ 
is manifestly of order $J_{ij}$.
The implicit parameter of this 
second approximation is evidently ``$J/h_0$''
where $J$ is an appropriately weighted average of the couplings, 
and $h_0$ is the averaged local field on (if that is not already 
uniform).  Then $h_0 \sim  J z$, where $z$ 
is an appropriately weighted count of the nearest neighbors. 
Thus, we are implicitly making {\it large coordination number}
approximation. It should work best on the fcc lattice ($z=12$),
or in lattices with $J_2$ comparable to $J_1$.
One expects it to be poorer quantitatively in (say) the
nearest-neighbor Kagom\'e lattice ($z=4$).  However, 
a biquadratic form with a fitted coefficient might
still be a reasonable approximation of the actual
functional form in that case.

Expanding the square in \eqr{eq:spwave-result} 
gives a constant term, a renormalization of the coupling due to
fluctuation so (in terms of the unit spins $\{ \nn _i \}$)
  \be 
     J^{\rm unit}_{ij} = J _{ij} S^2 - \frac{S}{4h_0} J_{ij}^2
   \eeq
and most importantly a term of the same {\it biquadratic} form
as introduced for thermal and dilution selection,
Eq.~\eqr{eq:biquad} with 
   \be
            K_{ij} \to \frac{S J_{ij}^2}{4 h_0}.
   \label{eq:quantum-Kij-final}
   \eeq
Because of this form quantum fluctuations
must favor collinear states.  This conclusion extends to 
all degenerate vector magnets with site-independent $h_0$.

\subsection{Comparison to alternate approaches}

As mentioned above, in the case of independent sublattices,
one obtains an effective Hamiltonian by perturbation in 
$J_1/J_2$, where $J_1$ is the inter-sublattice coupling. 
In the case of the $J_1$-$J-2$ square lattice antiferromagnet,
the spin-wave dispersion is easy enough to get analytically,
namely~\cite{FN-ODL}.
   \bea
     \hbar \omega(\qq) &=& S \Bigg\{
     \Big(C(\qq) -2 J_1 [\cos q_x+\cos q_y] \Big) \\
     \times \Big(C(\qq) &-& 2 J_1 \cos \Phi [\cos q_x-\cos q_y] \Big)
      \Bigg\}^{1/2}
   \eea
where $C(\qq)\equiv 4 J_2 (1-\cos q_x \cos q_y)$.
Integrating the zero-point energy over the zone 
and expanding to second order, we get
the spin wave energy (per unit cell)
   \be
       E_{\rm sw} =   E_{\rm sw}^{(0)} - 
            \frac{J_1^2}{8 J_2} G_{-1}  \cos^2\Phi
   \eeq
where
   \be
      G_{-1} \equiv \int \frac{d^2\qq}{(2\pi)^2}
           \frac{(\cos q_x-\cos q_y)^2}
               {1-\cos q_x \cos q_y}   \approx 0.727 .
   \eeq
If we match this with \eqref{eq:biquad} (with two bonds
per unit cell), noting $\cos^2\Phi= \big(\nn_i\cdot \nn_j\big)^2$
for nearest neighbors,   and remembering the local field is 
$h_0/S = 4 J_2$, 
we get 
    \be 
        K_{ij} = \frac{J_1^2}{4 h_0/S} G_{-1}.
    \eeq
Comparing to \eqr{eq:quantum-Kij-final}, we see
the real-space hopping approach overestimated
the biquadratic selection  energy by $\sim 33\%$,
which is not bad in view of the crudeness of the
approximation (and considering that the coordination
number is not so large  for this lattice.)

Two other tricks have been used, instead of the
real-space expansion trick, to arrive analytically 
at an simple effective Hamiltonian of biquadratic form
The first trick is expanding around an  infinite-range 
model~\cite{Zhang02}.
Comparing to that helps illuminate the derivation in this paper 
(Sec.~\ref{sec:quantum-real}, above), since
either approach is a way of setting up mean-field approximation.
That means breaking up the Hamiltonian as $\HH_0+\HH'$
such that $\HH_0$ has a trivial (and tractable) state that approximates 
the ordered state of interest, and expanding in $\HH'$.

Just as in statistical mechanics, there are two standard 
ways to set up the mean-field theory in real space.
In  the approach of Sec.~\ref{sec:quantum-real}, 
$\HH_0$ consisted of local fields 
that fix each spin in its classical direction, and the
wavefunction is a direct product of spin coherent states.
On the other hand, 
in Ref.~\onlinecite{Zhang02}, $\HH_0$ consists of infinite-range 
spin-spin couplings $J_{ij}$; they depend on the sublattice of 
$i$ and $j$, but are unchanged if either site is translated by any 
number of (magnetic cell) lattice vectors.  The latter approach, then,
is limited to states with a known kind of long-range order, 
such as face-centered cubic antiferromagnets, and cannot 
be used for highly frustrated antiferromagnets,  which have
a huge ensemble of non-periodic classical ground states.  
(Strictly speaking, the infinite-range approach is not even 
well posed if this ordered state gets modified into a texture with 
slowly varying directions, or if it has thermal spinwave fluctuations.
Thus, the advantage of the local-fields formulation is that it
handles arbitrary configurations of classical directions $\{ \nn_i \}$.
On the other hand, an advantage of the infinite-range 
formulation~\cite{Zhang02} is that it explicitly obtains
finite-size corrections in  the spin-wave selection energies.
(The corrections are big at system sizes tractable by exact diagonalization.)
In the systems where both approaches apply, they give the
same answer~\cite{Zhang02}.

A second trick works in the case of a pyrochlore lattice 
(or any other lattice made of corner sharing triangles and tetrahedra).
The sum in \eqr{eq:Eharm} can be represented
as the trace of the square root of the dynamical
matrix (basically $\HH'$).   A Taylor expansion of
that square root gave a series of terms which depend
on the classical directions of each spin.~\cite{Hen06,Hizi06} 

That was only applied to collinear ground states, but
(see Sec.~V of Ref.~\onlinecite{Hizi06}) it works
with small modifications for any ground state configuration.
It seems that the second order term of that expansion is
identical to the result we found here.  (In the collinear
case, this term was trivial, and nontrivial effects
were found only in higher-order terms representing
loops in the lattice.)  The trace viewpoint, however,
is more powerful in that it allows computation of
higher terms; also, in  some cases 
partial resummations of higher
terms might give better numerical approximations for
the coefficient in the effective Hamiltonian, 
analogous to the resummations in Ref.~\onlinecite{Hizi06}.

\subsection{Kagom\'e and similar antiferromagnets}

For some frustrated magnets, the effective Hamiltonian
\eqr{eq:spwave-result}
takes the same value in every classical ground state and
hence does not break the degeneracy.  In particular, this
happens for lattices (such as the kagom\'e) built from
triangles, each of which has a $120^\circ$ spin arrangement
in a ground state.  Since $\nn_i\cdot\nn_j=-1/2$ for all
nearest neighbor pairs, Eq.~\eqr{eq:spwave-result}
reduces to a constant.

In such systems, {\it coplanar} states are the closest
thing to collinear states within the ground state manifold,
and these indeed have the lowest harmonic zero-point energy.

\LATER{The next should be cut, I think.}
What would be the natural terms in an effective Hamiltonian,
analogous to \eqr{eq:spwave-result} and \eqr{eq:biquad}?  
Since three spins are required to distinguish a coplanar
from a noncoplanar state, one might think a term is needed
of form
   \bea
    |\nn_i \cdot \nn_j \times \nn_k|^2 &=& 1- 
    (\nn_i\cdot \nn_j)^2 - (\nn_j\cdot \nn_k)^2 - (\nn_k\cdot \nn_i)^2 
\nonumber \\
     &+& 2 (\nn_i\cdot \nn_j)(\nn_j\cdot \nn_k)(\nn_k\cdot \nn_i) .
   \label{eq:coplanar-triple}
   \eea

The simplest spin term which selects for the coplanar states 
is just a {\it second-neighbor} biquadratic term of
form Eq.~\eqr{eq:biquad}.
(In a nearest-neighbor Heisenberg model,
that would not get generated by \eqr{eq:spwave-result}.)
On the kagom\'e lattice
(and other lattices of corner-sharing triangles),
the third-neighbor coupling $K_3$ (for spins in the same row,
not across a hexagon) must be equal to the second-neighbor 
term $K_2$, for two reasons:
\begin{itemize}
   \item[] (i) if it were not, this would
  split the degeneracy between different {\it coplanar}
  states, whereas we know they are exactly degenerate
according to the harmonic-order selection energy
   \item[] (ii) at the perturbation order that this term mostly
comes from, there is no distinction between paths to
these two neighbor-of-neighbor sites.
    \end{itemize}

An extension of the Sec.~\ref{sec:quantum-real}
calculation to fourth order perturbation theory
does generate the desired neighbor-of-neighbor biquadratic
terms on the kagom\'e lattice.
(The trace form from Sec. V of Ref.~\onlinecite{Hizi06} is
probably an easier way to set this up.)
Specifically, $K_2$ and $K_3$ come from processes involving 
a string of sites $i$, $j$, and $k$ (not on the same triangle).  
There is a spin exchange involving the pair $(ij)$, another 
involving $(jk)$, and then the same two pairs (in either order) 
so that we return to the ground state of the trivial Hamiltonian 
$\HH_0$~\cite{FN-4thO}.

The phenomenological term adopted in Ref.~\onlinecite{vondelft}
is the same term that would follow from the second-neighbor
biquadratic interactions.  (This was assumed as 
the barrier potential to estimate a spin-tunneling 
amplitude in the kagom\'e antiferromagnet~\cite{vondelft}.)

Unfortunately, this functional form is wrong for the
Kagom\'e case~\cite{Ritchey}, and for coplanar states in general.
Instead, the effective Hamiltonian scales non-analytically
as $|\sin \theta|$, where $\theta$ is the angle between 
the spin planes in adjoining parts of the lattice~\cite{Hizi06}.
Sec. V of Ref.~\onlinecite{Hizi06} has clarified why 
the spin-wave energy cost, when spins rotate out of one
of the discretely  selected ground states, scales
linear in angle deviations for the (coplanar) Kagom\'e case 
but quadratic for the (collinear) pyrochlore case.
This cusp should get somewhat rounded once {\it anharmonic}
fluctuations are taken into account~\cite{vondelft}.
It is an interesting, unsolved challenge how to formulate
the calculation of a local effective Hamiltonian, 
in the spirit of \eqr{eq:spwave-result},
that captures the cusp behavior.

Nevertheless, a phenomenological form favoring coplanarity, 
such as a neighbors-of-neighbors biquadratic term, would be 
an improvement over classical simulations that do not account 
at all for quantum effects.  
In static correlations, as can be accessed by Monte Carlo~\cite{chalker-kag},
one expected  effect would be a much more robust stabilization
of coplanar states (at much higher temperatures).
In the dynamics, as can be simulated by 
molecular dynamics~\cite{robert-canals},
the coplanarity term gives a finite frequency to the band of 
zero-frequency spin wave modes (except the Goldstone mode).

\section{Discussion}
\label{sec:discussion}

To summarize, we have derived effective Hamiltonian terms
for selection free energies from spin wave fluctuations
due to thermal excitation,  quantum zero-point motion,  
or to dilution disorder;  the result has the form 
of an effective biquadratic Hamiltonian 
\eqr{eq:biquad}
in every case, with the coefficient proportional to temperature,
$1/S$, or dilution concentration, respectively.
In each of the three cases, our key trick was to
rework the problem until the spin configuration became 
the entries in a matrix, and then expanding in them.
The same idea has, in one guise or another, 
been successful for several independent examples of
getting an effective Hamiltonian by integrating out
fluctuations. Besides Ref.~\onlinecite{Henley1987}, 
which is a precursor (for the classical undiluted fcc lattice)
of our calculation in Sec.\ref{sec:classical}, this was
done for quantum~\cite{Chan94} or classical~\cite{Hen08-kacl}
kagom\'e systems, as well as 
pyrochlore antiferromagnets~\cite{Hizi06,Hizi-pyquart}.

Sometimes, it is possible to compute numerical energies for
a large database of configurations, and then fit an effective
Hamiltonian empirically~\cite{Hizi-pyquart}.
Even that brute-force approach depends critically on
analytics which suggest the proper functional form to
be used for fitting.

Now we turn to the applications of effective Hamiltonians.
They are convenient for predicting the ground
state spin pattern,  when the Hamiltonian is complicated by too
many kinds of interactions, e.g. spin anisotropies of all
sorts, dipolar interactions, distant-neighbor couplings, 
external fields, or magnetoelastic couplings; also, 
defects, inhomogeneities, boundaries and domain walls.
Thus, they allow quick understanding of phase diagrams
which may have many parameters.
For example, phenomenological biquadratic term representing the quantum 
fluctuations was used~\cite {jacobs} to explain a plateau in 
the dependence on external field of the incommensurate ordering
wavevector of CsCuCl$_3$.  Another one-dimensional
model with a biquadratic term was studied in Ref.~\cite{kaplan09}.

Effective Hamiltonians have also been combined with semiclassical
tunneling theory to understand tunneling processes,
when degenerate~\cite{Hen98,Houle99,Zhang02}
or highly degenerate~\cite{vondelft} ground states 
are separated by barriers.

\subsection{Pitfalls of classical simulations}

As one of us noted in Ref.~\onlinecite{Henley-HFM2000},
large $S$ justifies visualizing each spin as a
fixed-length vector, but it does {\it not} justify
a purely classical simulation of the system, 
as is commonly 
done~\cite{giebultowicz-sim,chalker-kag,huse,Reimersmc,moessner,robert-canals,zhitomirsky}.
Notwithstanding the well-defined spin directions,
the unmodified classical Heisenberg model
gives a {\it qualitatively wrong} picture of the behavior.
\LATER{Cite other old simulations? In particular, 
Reimers did both Kagome and pyrochlore.}

The reason can be expressed in terms of different energy scales,
as outlined in Ref.~\onlinecite{Henley-pyrice}.  
The highest scale is $E_J ~\sim JzS^2$, the scale of the mean-field
ordering temperature and Curie-Weiss constant; when $T\ll E_J $
it is already a good approximation that the system is in a classical
ground state with small fluctuations. 
Another scale is the temperature $T^*$ where ordering, freezing, or
other phase transitions take place:  by definition, a ``highly
frustrated'' magnet~\cite{ramirez} is one in which $T^*/ T_{\rm MF\ll 1}$;
the ratio can approach $10^{-2}$.
We are usually interested in lower temperatures $T\sim T^*$, 
where the interesting changes occur.

The scale which has been ignored is that of harmonic spin-wave fluctuations, 
$E_{\rm coll} \sim JS \lesssim E_J$.  As long as $E_{\rm coll} < T < E_J$, 
the system is certainly classical.  But we're most interested
in  $T\sim T^* \ll E_{\rm coll}$, where a classical description is
{\it not} valid.  Consider for example~\cite{Henley-HFM2000} the energy barriers
against flipping from one collinear (or coplanar) state to another:
those due to spin-wave zero-point energy are greater 
(by the ratio $E_{\rm coll}/T$) than the classical barriers. 
That will drastically affect any thermally activated dynamics, and
will enhance the ordering tendency (e.g. a large $S$ pyrochlore
antiferromagnet will develop a long range order at some $T>0$.).
Indeed, in the $T < E_{\rm coll}$ regime where the classical fluctuations 
are just a small correction to the quantum fluctuations, the $T$-dependent
selection free energy does not have the form it does in a purely
classical model~\cite{Sheng}.

I believe there is an easy fix: if the phenonmenological term 
\eqr{eq:biquad} (or whatever is appropriate for the system in
question) is added to the classical Hamiltonian by hand, the
modified classical simulation {\it can} closely 
represent the low-temperature behavior.  Of course, such 
a simulation are much tractable (in terms of temperatures,
system sizes, and observables) than quantum Monte Carlo
This could be incorporated both into Metropolis Monte Carlo (MC),
for modeling the thermodynamics and phase transitions, as
well as molecular dynamics, for modeling spin-wave excitations
or magnetic relaxation behaviors.   (If quantitative results
are needed, it would be best to empirically fit the
coefficients in \eqr{eq:biquad} first.) Occasionally a more
exact selection free energy has been used in MC simulations~\cite{Tch},
but that is enormously cumbersome (requires diagonalizing an $O(N)$
dimensional dynamical matrix every MC step.)  Having the form of
a local effective Hamiltonian makes it very easy to allow
for the quantum effects, which is necessary if the simulation
means to represent a Heisenberg magnet realistically.

\acknowledgments

This work was supported by NSF grant DMR-DMR-0552461
and its predecessors.
We thank Assa Auerbach, Evgenii Shender,
Uzi Hizi, Roderich Meissner, and Tom Kaplan
for helpful conversations.
The original work was completed at Boston University.

\appendix

\section{Alternative Method of Dilution Selection}
\label{app:alt-dilution}

We explore an alternative method of calculating the dilution 
selection energy $E\dil$, using Lagrange multipliers to enforce
the unit length constraints.  
Rather than the model dilution employed in Sec.\ref{sec:classical}, we consider
the more realistic case
\be
    J_{ij}\rightarrow J_{ij}\epsilon_i\epsilon_j \equiv
  p^2J_{ij} + \eta_{ij}J_{ij}, 
   \label{eq:appdefj} 
\eeq
where 
$\eta_{ij}\equiv \epsilon_i\epsilon_j - p^2$ for $(i\neq
j)$.  Here $\epsilon_i=1$ with probability $p$ and is 
otherwise zero.  Thus $\langle \eta_{ij} \rangle = 0$.  
Then,
\be
    \delta H\dil=p^2\delta H + \delta H^{\prime}, 
   \label{eq:apdilH} 
\eeq
where $\delta H= \half \sum_{i,j}A_{ij}\ssigma_i\cdot {\ssigma}_j$, and 
\be
    \delta H^{\prime}= E_0 - \half Nh_0p(1-p) + \delta H\REF
   + \delta H_{1}. 
   \label{eq:apdeldil} 
\eeq
The random exchange field contribution is 
\be
    \delta H\REF = \half \sum_{ij}A_{ij}\eta_{ij}
   \SS^{(0)}_i\cdot\SS^{(0)}_j, 
\eeq
which averages to zero and is
ignored as before for computing selection, and
\be
    \delta H_{1}=\sum_{i} \ssigma_i\cdot \hh_i. 
   \label{eq:apdil} 
\eeq
Here 
\be
    \hh_i =\sum_j 
    A_{ij}\eta_{ij}\SS^{(0)}_j, 
   \label{eq:newdil}  
\eeq
for realistic dilution, or 
\be
    \hh_i=\sum_j A_{ij}\epsilon_j \SS^{(0)}_j, 
   \label{eq:olddil}  
\eeq
for the model dilution of \eqr{eq:modis}.
To minimize the total zero temperature energy with respect to the
spin deviations $\ssigma_i$ for a {\it fixed} configuration
of the quenched disorder variables $\eta_{ij}$, we need to
include a Lagrange multiplier $\lambda_i$ to ensure that the unit length
constraints $\ssigma_i\cdot\SS^{(0)}_i=0$ are obeyed,
yielding 
\be
    p^2\ssigma_i=-(A^{-1})_{ij}(\hh_j + 
   \lambda_j\SS^{(0)}_j). 
   \label{eq:apsolvesig} 
\eeq
Introducing a matrix notation as in Sec.\ref{sec:classical} 
the $\lambda_i$ are determined by 
the constraint equation to be 
\be
   {\boldsymbol{\lambda}}= - \BB^{-1}\PP\AA^{-1}\hh. 
   \label{eq:aplamsol}
\eeq
When \eqr{eq:aplamsol} is substituted into \eqr{eq:apsolvesig}, and the result
substituted into \eqr{eq:apdilH}, we obtain an expression for the
total energy in the presence of the specific configuration of
dilution variables:
\bea
   p^2\delta H + \delta H_1 \equiv E\dil =\qquad\qquad&&
   \nonumber\\
      \half p^{-2} 
(\hh^T\AA^{-1}\PP^T\BB^{-1}\PP\AA^{-1}\hh &-&
       \hh^T\AA^{-1}\hh)
   \label{eq:apedil} 
\eea
identical to \eqr{eq:DeltaF}.

Both terms take the form $\hh^T \MM \hh$, which must now be 
averaged over the disorder.  In components the average becomes
\be
    \sum_{\mu\nu}\sum_{ijkl}M_{i\mu, j\nu}
   S^{(0)}_{l\mu}S^{(0)}_{k\nu}A_{il}A_{jk}\langle
   \eta_{il}\eta_{jk}\rangle . 
   \label{eq:extra}  
\eeq
The average is 
\bea
   \langle \eta_{il}\eta_{jk}\rangle &=& p^3(1-p)(\delta_{ij} +
   \delta_{ik} + \delta_{lj} + \delta_{lk}) \nonumber \\
   &+& p^2(1-p)^2(\delta_{ij}\delta_{lk} + \delta_{ik}\delta_{lj}),
    \label{eq:apavdis}  
\eea
if we also replace the $\AA$'s in \eqr{eq:extra}
by $\JJ$'s to ensure that $i\neq l,j\neq k$.  
The terms proportional to $(1-p)$ represent effects of 
independent missing spins (defects), and will be seen to 
reproduce the form of $E\dil$ calculated in Sec.\ref{sec:classical}.  
The $(1-p)^2$ terms account for pairwise correlations 
between different defects.   Thus, 
\bea
    \langle \hh^T \MM\hh \rangle &=& p^3(1-p) \Tr\big[
h_0 \PP \JJ \MM \PP^T+ h_0 \PP \MM \JJ \PP^T +  
             \nonumber \\
&&\PP \JJ \MM \JJ \PP^T + h_{0}^2 \PP \MM \PP^T\big]
    + p^2(1-p)^2 \times 
       \nonumber \\
       \times \sum_{ij,\mu\nu} &&
       J_{ij} J_{ij} S^{(0)}_{i\nu} \big[S^{(0)}_{j\mu} M_{j\nu,i\mu} 
   + S^{(0)}_{i\mu}M_{j\nu,j\mu} \big]
   \label{eq:apavm} 
\eea

It is now straightforward to evaluate $E\dil$ by substituting 
\eqr{eq:apavm} in \eqr{eq:apedil}.  Note that the second term in
\eqr{eq:apavm} is smaller than the first by order $1/z$ where $z$ is
the coordination number ($z=12$ or $18$ for face-centered cubic,
depending whether $J_2$ is included.)
After performing simplifications we find the dilution energy 
   \bea
   E\dil&=&\half p(1-p) \big[ \Tr\BB^{-1} - Nh_0 \big]
   \nonumber \\
   &+& \half (1-p)^2 \Bigg[
   + \half (1-p)^2 \Bigg[ \sum_{ij,\mu\nu} J_{ij} J_{ij} S^{(0)}_{i\nu} 
   \times \nonumber \\
     \Big( S^{(0)}_{j\mu} && \big[{\AA^{-1}\PP^T\BB^{-1} \PP \AA^{-1}-
     \AA^{-1}}\big]_{j\nu,i\mu}  \\
   &+& S^{(0)}_{i\mu}
   \big[{\AA^{-1} \PP^T \BB^{-1} \PP\AA^{-1}-\AA^{-1}}\big]_{j\nu,j\mu}
   \Big) \Bigg]
   \label{eq:aptotal} 
\eea
(Some simplifications of the ${\cal O}(1-p)^2$ terms are
possible, but the result cannot be written in terms of
conventional matrix operations.)

Clearly the first term in \eqr{eq:aptotal}
is just the dilution part of \eqr{eq:fsel} of
Sec.\ref{sec:classical}
if $\langle \epsilon^2\rangle$ is identified with $(1-p)/p$
(after rescaling $J_{ij}\rightarrow J^{\prime}_{ij}=J_{ij}p^2$).  
As argued in Sec.\ref{sec:classical} $E\dil=0$ for any collinear state.  Here
this is clearest from examination of \eqr{eq:apsolvesig}.  In a
collinear state ($T=0$), all the $\SS^{(0)}_i$ may be taken along the
${\bf \hat{z}}$ axis.  Then clearly the only allowed directions
for $\ssigma_i$ fluctuations are in the $xy$-plane,
eliminating the second two terms from \eqr{eq:apsolvesig} and forcing
$\ssigma_i=0$ for all $i$.  Clearly then $E\dil=0$.  On
the other hand the form of $p^2\delta H + \delta H_1$ shows that
in a generic noncollinear case, where $\ssigma_i\neq 0$, the
dilution energy for {\it every} realization of the quenched
fluctuations will be reduced---the system takes advantage of the
fluctuation.  Thus for a generic noncollinear state,
we must have $E\dil<0$ strictly.  

Systems with an {\it exceptional} degeneracy may have $E\dil=0$
even for noncollinear states.  For example, the 2D XY square
lattice AF with $J_2<0$ and $J_1=0$.  Then the system forms two
{\it decoupled} simple antiferromagnetic sublattices.  The
coupling from the dilution then comes only from spins on the same
sublattice, which are collinear.  
Thus $E\dil=0$ even though the total spin
configuration is noncollinear.

It should be emphasized that the terms above are not
the only contributions of  ${\cal O}(1-p)^2$,
since neglected terms of order 
$\ssigma_{i}^3$ also contribute.  
Only in a case where some small parameter 
controls the size of the 
$\ssigma_i$ will the exhibited terms give the dominant
${\cal O}(1-p)^2$ contribution.  

\section{Effects of Magnetoelastic Coupling}
\label{app:magnetoel}

It is well-known that exchange striction, 
an {\it isotropic} effect of nonzero
magneto-elastic coupling, also leads to an effective biquadratic exchange.  
It is long known these favor collinear states, e.g. in MnO, 
and they play important roles in highly frustated spinels~\cite{Tch-distortion}.
To see how the biquadratic form arises
in this case, we add to the Hamiltonian the elastic energy
$H_{\rm el}$ and 
magneto-elastic coupling $H_{\rm me}$.  The elastic energy is 
\be
    H_{\rm el}=\half \int d^3x
\varepsilon_{\kappa\lambda}\varepsilon_{\mu\nu}~~. 
   \label{eq:elen} 
\eeq
Here $\varepsilon_{\kappa\lambda}$ is the strain tensor, and 
   $ \{ c_{\kappa\lambda\mu\nu} \}$
are the elastic constants, separately symmetric in 
$(\kappa\lambda)$ and $(\mu\nu)$,
and under their interchange.  
(We use the summation convention for component indices.)  
The magneto-elastic coupling is given by 
\be
    H_{\rm me}=\half \sum_{\ix\jy}\varepsilon_{\kappa\lambda}
\kappa\lambda
   \alpha^{\kappa\lambda}_{\ix\jy}\SS_\ix\cdot\SS_\jy~~, 
   \label{eq:meen} 
\eeq
obtained simply by expanding the exchange constants in lattice
displacements. 
Thus the magneto-elastic constants are defined by 
\be
    \alpha^{\kappa\lambda}_{\ix\jy}= -\big(\rr_{ij}\big)_\kappa
\frac{\partial_l J_{\ix\jy}}{\partial \big(\rr_{ij}\big)_\lambda}
   \label{eq:defme} 
\eeq
Next the energy $H_{\rm el}+H_{\rm me}$ is minimized with respect to the
strains $\varepsilon_{kl}$ and the result substituted back in 
this expression.  The result is
\be
    H^{eff}_{\rm me}=-\frac{1}{8}\sum_{\ix\jy}\sum_{\kz\lw}
   [c^{-1}]_{\kappa\lambda\mu\nu} 
   \alpha^{\kappa\lambda}_{\ix\jy}
    \alpha^{\mu\nu}_{\kz\lw}
   (\SS_\ix \cdot \SS_\jy) (\SS_\kz \cdot \SS_\lw)~~. 
   \label{eq:mefinal} 
\eeq
This has the biquadratic form.  Within the ground state manifold
\eqr{eq:mefinal} has an expression in terms of the same quartic invariants 
mentioned at the end of Sec.~\ref{sec:effham}.

To estimate the size of this effect, we assume the exchange
constants decay rapidly with distance, according to 
$J(r)=J_0(r/r_0)^{-f}$,~\cite{super} 
yielding 
\be
    \alpha^{\kappa\lambda}(\rr)
   \approx - \frac{f}{2}  \frac{r_\kappa r_\lambda}{r^2} J(\rr)~~. 
   \label{eq:meapp}  
\eeq
There is some evidence that for superexchange, $f$ is between
10,~\cite{Bloch} and 14.~\cite{super}


\begin{thebibliography}{99}
\bibitem{fncu}
Eq. \eqr{eq:modis} is actually correct for Cu nuclear spin disorder, 
where the difference in the nuclear moments for different Cu
isotopes provides the disorder.
\bibitem{FN-PPnotsquare}
One can't simply change 
variables in \eqr{eq:partb} by a shift of 
$t_i$ in the integral, since  $\PP$ is not square.
\bibitem{fnweight}
Any weighting of the ground states that gives the same 
weight to symmetry-equivalent ground states would be acceptable, 
since $\Bbar$ would have all lattice symmetries restored.
\bibitem{FN-LT-bave}
When the classical ground state is built from the optimal eigenmodes 
of the $J_{ij}$ matrix, there is another simplification:
$\langle \SS_i\cdot \SS_j \rangle_s$ in \eqr{eq:bave} 
is simply proportional
to the autocorrelation of those eigenmodes, averaged over
the star of degenerate wavevectors.
\bibitem{FN-ODL}
In Ref.~\onlinecite{HenleyPRL}, the spin wave dispersion
used in the quatum selection calculation had  a different 
form, as a large easy-plane anisotropy $\Gamma$ was assumed there.
\bibitem{FN-4thO}
Besides biquadratic terms involving all pairs in
   \eqr{eq:coplanar-triple},
fourth-order perturbation also generates the last term in 
\eqr{eq:coplanar-triple}.  However, this has no selection
effect: since $\nn_i\cdot\nn_j=\nn_j\cdot\nn_k=-1/2$ in any
ground state, the term reduces to a sum of $\SS_i\cdot\SS_k$
terms, which in turn reduces to a constant (for any ground
state) after combining the equal contributions  from
both neighbor-of-neighbor separations.
\end{thebibliography}
\end{document}